\documentclass[12pt]{iopart}

\usepackage{iopams}
\usepackage{graphicx}
\usepackage{dsfont}
\usepackage{pxfonts}
\usepackage{color}

\makeatletter
\def\revddots{\mathinner{\mkern1mu\raise\p@
\vbox{\kern7\p@\hbox{.}}\mkern2mu
\raise4\p@\hbox{.}\mkern2mu\raise7\p@\hbox{.}\mkern1mu}}
\makeatother

\newcommand{\half}{\mbox{$\textstyle \frac{1}{2}$}}

\newcommand{\up}{\uparrow}
\newcommand{\dn}{\downarrow}

\begin{document}

\title[Fragile entanglement statistics]
{Fragile entanglement statistics}

\author{Dorje C. Brody$^{1,2}$, Lane P. Hughston$^{1,2}$, and David M. Meier$^1$}

\address{$^1$Department of Mathematics, Brunel University London, Uxbridge UB8 3PH, UK  \\ 
              $^2$Department of Optical Physics and Modern Natural Science, \\ 
St Petersburg National Research University of Information Technologies, Mechanics and Optics, 
49 Kronverksky Avenue, St Petersburg 197101, Russia}

\date{\today}

\begin{abstract}
If X and Y are independent, Y and Z are independent, and so are X and Z, one might 
be tempted to conclude that X, Y, and Z are independent. 
But it has long been known 
in classical probability theory that, intuitive as it may seem, this is not 
true in general. 
In quantum mechanics one can ask whether analogous statistics can
emerge for configurations of particles in certain types of entangled states. 
The 
explicit construction of such states, along with the specification of suitable sets of  
observables that have the purported statistical properties, is not 
entirely straightforward. 
We show that an example of such a configuration arises in the 
case of an $N$-particle GHZ state, and we are able to identify a family of observables 
with the property  that  the associated measurement outcomes are independent for any 
choice of $2, 3, \ldots,N-1$ of the particles, even though the measurement outcomes for 
all $N$ particles are not independent. 
Although such states are highly entangled, the 
entanglement turns out to be 'fragile', i.e.~the associated density matrix has the property 
that if one traces out the freedom associated with even a single particle, the resulting 
reduced density matrix is separable. 
\end{abstract}


\submitto{\JPA}


\section{Introduction} 

\noindent The notion of statistical dependence can at times be counterintuitive, leading to 
surprising conclusions even in classical probability \cite{stoyanov}. In the case 
of a quantum system consisting of two or more particles (or subsystems), the intrinsic 
quantum statistical dependence of the various constituents upon one another, otherwise 
known as `entanglement', can admit features---such as violation of the Bell inequality---that 
do not have obvious classical analogues. Indeed, ever since Schr\"odinger's introduction 
of the idea of `entanglement' \cite{schrodinger,schrodinger2}, and the concurrent work of 
Einstein, Podolsky and Rosen on its paradoxical aspects \cite{EPR}, a great deal of effort 
has been made by researchers to understand the implications of this fundamental notion, and to seek 
applications (see, e.g., articles in \cite{JPA} and references cited therein). 

The purpose of the present paper is to point out the peculiar statistics  arising from measurements of a certain set of observables when 
a  system of particles is prepared in a particular type of entangled state. Our considerations lead to 
a quantum counterpart of a surprising result in classical probability due to 
Bohlmann~\cite{bohlmann} and Bernstein~\cite{bernstein}: to wit, that it is possible to 
construct a situation in which while X and Y are independent, Y and Z are independent, 
and X and Z are independent, nevertheless X, Y, and Z are not independent. 
We show, more generally,  that when measurements are made on a quantum system consisting of $N$ spin-$\frac{1}{2}$ particles, a version of the Bernstein phenomenon arises when the state of the system belongs to a particular class of fully-entangled states that we shall refer to as  `Bernstein states'.
We construct 
this family of states explicitly and show that no other states give rise to Bernstein statistics. 
The elements of a special subclass of these states turn out to be equivalent, modulo local unitary transformations of the pure phase type, to $N$-particle 
GHZ states \cite{ghz}. While such GHZ-equivalent states are known to be strongly entangled, they are at the same time `fragile' in the sense that if one traces  over the contribution of any one of the 
constituent particles then the entanglement is destroyed---that is to say, the resulting 
reduced density matrix of the remaining $N-1$ particles is separable. 

The paper is organized as follows. In Section~\ref{sec:2} we review 
the classical Bernstein distribution, and construct an example of a three-particle quantum Bernstein 
state by analogy with the classical case. This state is shown to 
be fragile. In Section~\ref{sec:3}, the three-particle example introduced in Section~\ref{sec:2} is generalized 
by use of a combinatorial argument to the construction of
a special $N$-particle  Bernstein state  for each $N \geq 3$. 
In Section~\ref{sec:4}, we show by use of a basis transformation that the special $N$-particle  Bernstein states  are equivalent to 
a class of $N$-particle GHZ states and that these states are fragile. 
In Section~\ref{sec:5}, the $N$-particle Bernstein states
constructed in Section~\ref{sec:3}  are shown
to be unique in having the Bernstein distribution, up to the inclusion of a set of $2^{N-1}$ phase factors. 
In Section~\ref{sec:6} we show that for $N>3$ the space of Bernstein states that are equivalent modulo local phase transformations to the special Bernstein states, for which the phase factors take the value unity, is a proper subspace of the space of general Bernstein states. In the three-particle case, the special Bernstein states and the general Bernstein states are equivalent modulo local phase transformations. In 
Section~\ref{sec:7} we construct a state with an 
inhomogeneous Bernstein distribution and show that the fragility property holds  in the 
homogeneous limit. 
Finally, in Section~\ref{sec:8} we show how the so-called Mermin paradox \cite{mermin, peres} can be extended by the use of our methods to the case of an $N$-particle GHZ state.

\section{Bernstein statistics} 
\label{sec:2}

For the benefit of readers less well acquainted with the relevant ideas, first we introduce the 
Bernstein distribution~\cite{bernstein,Kolmogorov}. Consider a set of four cards, on which the 
following sequences of numbers are printed: $(110)$, $(101)$, $(011)$, and $(000)$. The 
four cards are placed in a bag, from which one card is selected at 
random. Let X be the event that the first digit on the selected card 
is 1, let Y be the event that the second digit on the selected card is 1, and let Z be the 
event that the third digit on the selected card is 1. Clearly, the probability of 
X is $\frac{1}{2}$, the probability of 
Y is $\frac{1}{2}$, and the probability of 
Z is $\frac{1}{2}$. Similarly we find that the probability that X and Y 
occur is $\frac{1}{4}$, the probability that X and Z 
occur is $\frac{1}{4}$, and the probability that Y and Z 
occur is $\frac{1}{4}$. Thus, the triplet (X, Y, Z) 
is pairwise independent. Now, if X, Y and Z were independent, the probability of all three events occurring would be 
$\frac{1}{8}$. Since the probability of all three
events occurring is zero, it follows that X, Y and Z are not independent. 

A quantum version of the 
 Bernstein phenomenon can be constructed as follows. We consider a system of three spin-$\frac{1}{2}$ particles.
Write $\sigma_{nx}$, 
$\sigma_{ny}$, $\sigma_{nz}$ for the spin operators associated with particle  $n$, 
where $n = 1, 2, 3$, and write $\pi^{\pm}_{nx}$, $\pi^{\pm}_{ny}$, $\pi^{\pm}_{nz}$ 
for the corresponding projection operators onto the associated eigenstates. Thus,  
$\pi^{\pm}_{nx} = \half (1\pm \sigma_{nx} )$,
$\pi^{\pm}_{ny} = \half (1\pm \sigma_{nz} )$,
and $\pi^{\pm}_{nz} = \half (1\pm \sigma_{nz} )$. In what follows, we work mainly with $\sigma_{nz}$ eigenstates.
Thus, for example, 
we write $| \, \!\!\uparrow\uparrow\uparrow
\rangle$ for the state vector of a system consisting of three spin-$\frac{1}{2}$ particles in
$\sigma_{1z}$, $\sigma_{2z}$, $\sigma_{3z}$ spin-up states. An explicit 
example of a Bernstein state in the case of a system of three spin-$\frac{1}{2}$ particles 
is 
\begin{eqnarray}
|B^{(3)}\rangle =\frac{1}{2}\big( \,
 |\! \uparrow\uparrow\downarrow\rangle + |\! \uparrow\downarrow\uparrow\rangle + 
|\! \downarrow\uparrow\uparrow\rangle + |\! \downarrow\downarrow\downarrow\rangle \, \big) \, . 
\label{eq:1} 
\end{eqnarray}
Evidently, $|B^{(3)}\rangle$ is neither a product state nor a partial product state, so it 
represents a totally entangled state.  The structure of the expression on the right side of 
(\ref{eq:1}) allows one to verify the Bernstein property for spin measurements in the $z$ direction. To this end,  let us write $(+\bullet\bullet)$ for the event that the outcome of a 
measurement of ${\sigma}_{1z}$ on the first particle is spin up; similarly, let us write 
$(++\bullet)$ for the event that the outcomes of measurements of ${\sigma}_{1z}$ and 
${\sigma}_{2z}$ on the first and second particles, respectively, are both spin up, and so on 
for the various other possible outcomes for single, double, and triple spin measurements. 
One sees that when the system is in the state $|B^{(3)}\rangle$ the 
probabilities of the outcomes of the various single spin measurements are given by
\begin{eqnarray} 
{\mathbb P}(+\bullet\bullet)={\mathbb P}(\bullet+\bullet)={\mathbb P}(\bullet\bullet+)
=\frac{1}{2},
\label{eq:2}
\end{eqnarray}  
along with similar results for spin down. Then for the probabilities of the outcomes 
of spin measurements on any pair of the three particles, we have
\begin{eqnarray} 
{\mathbb P}(++\bullet)={\mathbb P}(+\bullet+)={\mathbb P}(\bullet++)=\frac{1}{4},
\label{eq:3}
\end{eqnarray}
with similar results if any `up' is replaced with a `down'. In particular, we find that  
\begin{eqnarray}
{\mathbb P}(++\bullet)={\mathbb P}(+\bullet\bullet)\times 
{\mathbb P}(\bullet+\bullet),
\end{eqnarray}
and that similar decompositions into products of single spin measurement outcome 
probabilities hold for all possible outcomes of the three pairs. It follows that the outcomes 
for $\sigma_{1z}$, $\sigma_{2z}$, and $\sigma_{3z}$ are statistically independent for any 
pair of these measurements. On the other hand, if one measures the spins of all three 
particles the situation changes. In particular, we find that the probability of three `ups' is given by
\begin{eqnarray} 
{\mathbb P}(+++)= 0 \neq 
{\mathbb P}(+\bullet\bullet)\times{\mathbb P}(\bullet+\bullet)\times{\mathbb P}(\bullet\bullet+)
=\frac{1}{8},
\label{eq:4}
\end{eqnarray}
which shows that the outcomes of the three spin  measurements are 
not independent even though they are pairwise independent. In conclusion, one sees that when a quantum system is 
prepared in the state $|B^{(3)}\rangle$, the statistics of the outcomes of measurements of the three spin operators 
$\sigma_{1z}$, $\sigma_{2z}$, and $\sigma_{3z}$  are those of Bernstein.

The Bernstein state  (\ref{eq:1})  is remarkable in another respect as well: the associated pure density matrix 
$|B^{(3)}\rangle \,  \langle B^{(3)} |$ lies in the subclass of three-particle states having the property that if one traces out the degrees of freedom belonging to any one of the three particles then the resulting reduced density matrix is 
separable. This is the so-called  `fragility' property that we alluded
to earlier.  To verify that 
$|B^{(3)}\rangle \,  \langle B^{(3)} |$ is fragile, we  take the trace over, say, the first 
particle ${\rm P}_1$, to obtain the reduced density matrix ${\rho_{23}}={\rm tr}_{{\rm P}_1} 
\, |B^{(3)}\rangle\langle B^{(3)}|$ associated with the system consisting of the remaining two particles. 
Note that if, in the original three-particle state, one were to measure the first spin in the $z$ direction, then conditional on 
an `up' outcome the state of the remaining two particles would be 
$\frac{1}{\sqrt{2}}\, ( \, |\!\!\uparrow\downarrow\rangle + 
|\!\!\downarrow\uparrow\rangle \, )$. 
Likewise, if the outcome of the measurement of the first spin is `down', the state 
of the remaining two particles would be $\frac{1}{\sqrt{2}} \, (|\!\!\uparrow\uparrow\rangle + 
|\!\!\downarrow\downarrow\rangle)$. In both cases, the resulting two-particle state is maximally entangled
according to the standard geometric measure of entanglement \cite{BH}. 
On the other hand, if we trace out the first particle, the reduced density matrix of the two-particle system that results takes the form
\begin{eqnarray}
{\rho_{23}} = \frac{1}{4} \,
\left( \begin{array}{cccc} 1 & 0 & 0 & 1 \\ 0 & 1 & 1 & 0 \\ 0 & 1 & 1 & 0 \\ 1 & 0 & 0 & 1 
\end{array} \right)
\label{rho23}
\end{eqnarray}
when it is expressed in the basis 
$( \, |\!\!\up\up\rangle, |\!\!\up\dn\rangle, |\!\!\dn\up\rangle, |\!\!\dn\dn\rangle \, )$. The fact that this state is separable---in other words, that it can be represented as a mixture of disentangled states---can be verified by use of 
the Peres--Horodecki criterion \cite{Peres,Horodecki3}, which says that ${\rho_{23}}$ is separable if the partial transpose of ${\rho_{23}}$ has 
nonnegative eigenvalues. But it is clear that ${\rho_{23}}$ is invariant under partial 
transposition of either subsystem, and it has the nonnegative eigenvalues $\frac{1}{2}$ and 
$0$, each with multiplicity two. It follows that the reduced density matrix is separable. 
Indeed, it is a remarkable fact that a mixture of two maximally entangled pure states can 
nevertheless be separable. This can be seen somewhat more directly by use of the identity
\begin{eqnarray}
{\rho_{23}} &=&  
 \frac{1}{2}  \, ( \, | \! \!\uparrow\uparrow \rangle + 
|\!\!\downarrow\downarrow \rangle \, ) \,\,
( \, \langle \uparrow\uparrow \! | + 
\langle \downarrow\downarrow \! | \, )
+
 \frac{1}{2}  \, ( \, | \! \uparrow\downarrow\rangle + 
|\!\downarrow\uparrow\rangle \, ) \,\,
( \, \langle \uparrow\downarrow \! | + 
\langle \downarrow\uparrow \! | \, )
\nonumber \\ 
&=& 
\frac{1}{2}  \, | \! \! \leftarrow\leftarrow\rangle \, 
\langle \leftarrow\leftarrow \! | \,
+
 \frac{1}{2}  \,  | \! \rightarrow\rightarrow \rangle \, 
\langle \!  \rightarrow\rightarrow \! | \, ,
\label{disentangled}
\end{eqnarray}
where $| \!\!\leftarrow \rangle =  \frac{1}{\sqrt{2}}\,  (\,  |\!\!\up\rangle + |\!\!\dn\rangle ) $ and 
$|\!\rightarrow \rangle =  \frac{1}{\sqrt{2}}\, (\,  |\!\!\up\rangle - |\!\!\dn\rangle ) $, which expresses the reduced state $\rho_{23}$ 
as a mixture of entangled states and also as a mixture of disentangled states.  

Proceeding analogously, it should be evident then that similar results to those described 
above are obtained if we trace out the second particle ${\rm P}_2$ or the third particle 
${\rm P}_3$. From this, we conclude that the Bernstein state $|B^{(3)}\rangle$ defined by (\ref{eq:1}) 
is indeed a fragile state in the sense that we have indicated.

A useful consequence of (\ref{disentangled}) is that the independence of the measurement outcomes  of $\sigma_{2z}$ and $\sigma_{3z}$ can be seen to follow directly as a consequence of the symmetry of
($\ref{disentangled}$) with respect to the left and right arrow states. One can verify, for example, that 
${\rm tr} \, \pi^+_{2z} \, \rho_{23} =  \frac{1}{2}$, 
${\rm tr} \, \pi^+_{3z} \, \rho_{23} =  \frac{1}{2}$, and
${\rm tr} \, \pi^+_{2z} \,\pi^+_{3z} \, \rho_{23} =  \frac{1}{4}$, 
which are the correct probability relations for independence of the indicator functions for the 
second particle having spin up and the third particle having spin up. Here we have used the fact that when measurements are made only on particles 2 and 3, it suffices to make calculations using the reduced state ${\rho_{23}}$. 
On the other hand, it should be noted that the separability of a two-particle density matrix such as ${\rho_{23}}$ does not imply that the 
measurement outcomes of {\it all} bipartite observables are statistically independent. That can only happen when both the density matrix and the relevant observables factorize. In 
particular, the reduced state $ \rho_{23}$, which is separable, but not factorizable, gives 
rise to correlations. For example, we have
${\rm tr} \, \pi^+_{2x} \, \rho_{23} =  \frac{1}{2}$, 
${\rm tr} \, \pi^+_{3x} \, \rho_{23} =  \frac{1}{2}$, and
${\rm tr} \, \pi^+_{2x} \,\pi^+_{3x} \, \rho_{23} =  \frac{1}{2} \neq  
{\rm tr} \, \pi^+_{2x} \, \rho_{23} \,\, {\rm tr} \, \pi^+_{3x}  \rho_{23} = \frac{1}{4} $.
Hence, the outcomes of $\sigma_{2x}$ and $\sigma_{3x}$ are positively correlated, but not the outcomes of $\sigma_{2z}$ and $\sigma_{3z}$, which are 
independent. 

\section{$N$-particle Bernstein states} 
\label{sec:3}

The example considered in the previous section can be generalized in various ways. First, we can ask whether the state (\ref{eq:1}) is in some appropriate sense unique in admitting the Bernstein distribution for the indicated measurement outcomes. 
One possible generalization would be to consider a situation where the spin measurements on the various particles are taken with respect to a different choice of spin axis for each particle. But this set-up is equivalent to the original one in the sense that the outcome statistics are preserved if we make local unitary transformations on each of the spin operators in such a way as to line up the axes in the direction of a single choice of $z$-axis, and if at the same time we make the corresponding conjugate local unitary transformations on the various particles composing the overall state of the system. Thus without loss of generality  we may assume, as we have done in the previous section, that the axes of the various spin operators are aligned along a fixed given axis, which we have taken to be the $z$-axis. 

On the other hand, even if we stick with the given set of spin measurements we note that the Bernstein distribution still 
arises if we generalize the state by inserting unimodular phase factors in front of each term appearing in (\ref{eq:1}). Before exploring the ramifications of the resulting degrees of freedom in this apparently larger family of three-particle Bernstein states, it will be convenient to consider a further generalization of the Bernstein state (\ref{eq:1}), namely, that arising when we 
consider a system of $N$ spin-$\frac{1}{2}$ particles in a state that is suitably analogous to (\ref{eq:1}), 
i.e.~for which all of the terms have equal coefficients and for which the associated spin measurements admit an appropriate generalization of the Bernstein distribution.  Having constructed the $N$-particle analogues of (\ref{eq:1}), we turn our attention to the broader class of $N$-particle Bernstein states for which the phase degrees of freedom are included. 

The generalization of Section \ref{sec:2} to the four-particle case is not entirely obvious, 
since for $N=4$ we wish to construct a four-particle entangled state and an appropriate 
associated set of observables such that the resulting probability distribution exhibits not only pairwise 
independence but also triplet-wise independence. 
Various intermediate cases 
can also be considered. 
For example, in the case of a four-particle system, one might 
consider the construction of a state for which pairwise measurements were independent, but triplet-wise
and quadruplet-wise measurements were not. In what follows, we consider the situation where correlations emerge only at the highest level of participation. 

The fragility property likewise can be understood  as holding in a suitably hierarchical sense. For $N=4$, the highest degree of fragility arises 
in the case of a four-particle state with the property that the associated three-particle reduced  states are separable. 
But one can also consider the construction of entangled states with the property that the associated 
three-particle reduced  states are not separable, and such that entanglement is destroyed only 
at the level of the two-particle reduced states. The entanglement of such states 
is thus in some sense more robust. In what follows, we shall be mainly concerned with fragility of the highest degree. 

An example of a 
four-particle Bernstein state $|B^{(4)}\rangle $ that can be regarded as suitable generalization of 
the three-particle state $|B^{(3)}\rangle $ is given as follows:
\begin{eqnarray}
|B^{(4)}\rangle &=& \frac{1}{2\sqrt{2}}\Big( |\!\up \up \up \dn \rangle + |\!\up \up \dn \up \rangle 
+ |\!\up \dn \up \up \rangle + |\!\dn \up \up \up \rangle \nonumber \\ && \quad~\,+ |\! \up \dn \dn \dn 
\rangle + |\!\dn \up \dn \dn \rangle+ |\!\dn \dn \up \dn \rangle+|\!\dn \dn \dn \up \rangle \Big)\,  . 
\label{eq:6}
\end{eqnarray} 
It can be verified that if a ${\sigma}_z$ measurement is made on any one of the four particles, then the probabilities of the `up' outcomes are 
\begin{eqnarray} 
{\mathbb P}(+ \bullet \bullet \,\bullet)={\mathbb P}(\bullet+\bullet\,\bullet)={\mathbb P}(\bullet\bullet+\, \bullet)
={\mathbb P}(\bullet\bullet\bullet\, +)=\frac{1}{2}.
\label{eq:7}
\end{eqnarray} 
Then we find that for the probabilities of two of the spins being `up' we have
\begin{eqnarray} 
\fl \, \, \, {\mathbb P}(++\bullet\, \bullet)={\mathbb P}(+\bullet+\, \bullet)={\mathbb P}(+\bullet\bullet\, +)=
{\mathbb P}(\bullet++ \, \bullet)={\mathbb P}(\bullet+\bullet\, +)={\mathbb P}(\bullet\bullet+ \, +)
=\frac{1}{4} ,
\label{eq:8}
\end{eqnarray}
with similar results for the various other possible outcome pairs, thus allowing us to conclude that the outcomes of the spin measurements are 
pairwise independent. Likewise, for the probabilities of three of the spins being `up' we find that
\begin{eqnarray} 
{\mathbb P}(+++ \, \bullet)={\mathbb P}(++\bullet \, +)={\mathbb P}(+\bullet+ \, +)=
{\mathbb P}(\bullet++ \, +)=\frac{1}{8} ,
\label{eq:9}
\end{eqnarray}
with similar results for the various other possible outcome triples, thus showing that the 
outcomes of the spin measurements are triplet-wise independent. Yet, when we make ${\sigma}_z$ measurements on all four particles we find that the outcomes are not independent. In particular, we have 
\begin{eqnarray} 
\fl \quad \quad
{\mathbb P}(+++\, +)= 0 \neq 
{\mathbb P}(+\bullet\bullet \, \bullet)\times{\mathbb P}(\bullet+\bullet \, \bullet)\times 
{\mathbb P}(\bullet\bullet+ \, \bullet)
\times{\mathbb P}(\bullet\bullet\bullet \, +) = \frac{1}{16}. 
\label{eq:10}
\end{eqnarray}
One sees, therefore, that when a quantum system is 
prepared in the state $|B^{(4)}\rangle$, the outcomes of measurements of the four spin operators 
$\sigma_{1z}$, $\sigma_{2z}$, $\sigma_{3z}$, and $\sigma_{4z}$ have Bernstein statistics in an appropriately generalized sense.
One can check, moreover, that the various three-particle reduced density matrices, for example, 
${\rho}_{234}={\rm tr}_{{\rm P}_1}(|B^{(4)}\rangle\langle B^{(4)}|)$, are separable---the proof for 
general $N$ is given below---and hence that the state $|B^{(4)}\rangle$ defined by (\ref{eq:6}) 
has the fragility property. 

The combinatorial considerations leading to the explicit form of the fragile Bernstein state $|B^{(4)}\rangle$ and, more generally, to the construction
of a special Bernstein state $|B^{(N)} \rangle$  for any number of particles, can be summarized as follows. 

First, since the probability of an `up' outcome from a spin measurement has to be 
$\frac{1}{2}$, for each particle the $|\!\!\up\rangle$ and $|\!\!\dn\rangle$ states must appear 
with an equal number. Here we assume, for now, that the squared amplitudes of the 
coefficients appearing in the expansion of the state are all equal, as
in (\ref{eq:1}) and (\ref{eq:6}). Later we shall prove that this is a
necessary condition for a Bernstein state.  The 
number of terms in the expansion of $|B^{(N)}\rangle$ must therefore be even, say, $2k$ for some 
$k$. 

Second, since each pairwise `up' outcome has to have the probability $\frac{1}{4}$, 
the number of terms containing the pairwise-up combination 
$|\cdots\up\cdots\up\cdots\rangle$ has to be $\frac{1}{2}k$. Third, since the triplet positive 
outcome occurs with probability $\frac{1}{8}$, the number of terms of the form 
$|\cdots\up\cdots\up\cdots\up\cdots\rangle$ has to be $\frac{1}{4}k$. 

Continuing with this line of logic, we observe that the number of terms containing $n$ 
spin-up states must be $(\frac{1}{2})^{n-1} k$, and hence for $n=N-1$ we deduce that the 
number of terms containing $N-1$ spin-up states is $(\frac{1}{2})^{N-2}k$. But since the 
numbers of terms $\frac{1}{2}k$, $\frac{1}{4}k$, $\cdots$, $(\frac{1}{2})^{N-2}k$, have to 
be integers, 
we deduce that $k$ can only be $2^{N-2}$ or $2^{N-1}$. The latter case corresponds to 
an equal superposition of all the $2^N$ basis states for which the probability of $N$ 
spins pointing in the up direction is $2^{-N}$, not zero, which has to be ruled out; so we 
deduce that $k = 2^{N -2}$. This argument leads to the construction of the fragile Bernstein state
$|B^{(N)}\rangle$ for each $N\geq3$. 

As an example, let us show how one arrives at the state (\ref{eq:6}). We know 
that $|B^{(4)}\rangle$ is made up of $2k = 8$ terms. Four of them have the form 
$|\!\!\up \bullet \bullet \bullet \rangle$ and four of them have the the form 
$|\!\!\dn \bullet \bullet \bullet \rangle$. Of the first four, there are two of the form 
$|\!\! \up \up \bullet\,\bullet \rangle$ and the remaining two are of the form 
$|\!\!\up \dn \bullet\,\bullet \rangle$. One state in the former pair looks like 
$|\!\!\up \up \up \bullet \rangle$ and therefore must be $|\!\! \up \up \up \dn \rangle$ 
since the particles cannot all be found in up states. The other state in that pair 
has the form $|\!\!\up \up \dn \bullet \rangle$ and is therefore $|\!\!\up\up\dn\up\rangle$ 
since the configuration with the first, second and fourth particles in up states is present 
exactly once. Now consider the two states of the form 
$|\!\!\up \dn \bullet\,\bullet \rangle$. We see that one of them is of the type 
$|\!\!\up\dn\up\bullet\rangle$ and the other one is of the type  
$|\!\!\up \dn \dn \bullet\rangle$, which follows since there must be exactly two terms 
with the first and third particles in up states. The first of these types can be completed as 
$|\!\!\up \dn \up \up \rangle$, since the configuration with first, third and fourth particles in 
up states has to be present once. But this implies that the other type, 
$|\!\!\up\dn\dn\bullet\rangle$, can be completed as 
$|\!\!\up \dn \dn \dn \rangle$ since there can only be two terms that have the first and fourth 
particle in up states. The correct expression for the remaining four terms, of type  
$|\!\!\dn \bullet \bullet \bullet \rangle$, can be identified in a similar manner. This leads to   (\ref{eq:6}) if one takes the phase factors multiplying the various terms just identified each to be unity. 
The 
same kind of reasoning can be employed for general  $N$. In the case of  $N = 5$, for instance, one finds that the special Bernstein state takes the following form:
\begin{eqnarray}
|B^{(5)}\rangle &=& \frac{1}{4}\Big( |\!\up\up\up\up\dn \rangle + |\!\up\up\up\dn\up \rangle 
+ |\!\up\up\dn\up\up \rangle + |\!\up\dn\up\up\up \rangle \nonumber \\ && 
+ |\!\dn\up\up\up\up \rangle + |\!\up\up\dn\dn\dn \rangle + |\!\up\dn\up\dn\dn \rangle 
+ |\!\up\dn\dn\up\dn \rangle \nonumber \\ && 
+ |\!\up\dn\dn\dn\up \rangle + |\!\dn\up\up\dn\dn \rangle + |\!\dn\up\dn\up\dn \rangle 
+ |\!\dn\up\dn\dn\up \rangle  \nonumber \\ && 
+ |\!\dn\dn\up\up\dn \rangle + |\!\dn\dn\up\dn\up \rangle 
+ |\!\dn\dn\dn\up\up \rangle + |\!\dn\dn\dn\dn\dn \rangle \Big) \, . 
\label{eq:11}
\end{eqnarray} 
The fact that $|B^{(5)}\rangle$ exhibits the required Bernstein statistics when ${\sigma}_z$-up 
measurements are performed on each particle is ensured by the construction. 

\section{Relation to GHZ states} 
\label{sec:4} 

The special Bernstein state $|B^{(N)}\rangle$ thus constructed for each $N$ has the 
property that whenever one of the particles is traced out the resulting density matrix is 
separable. How to verify  the 
separability of the reduced density matrices is not immediately obvious on account of 
the lack of a general sufficient condition for separability. It turns out, nevertheless, that these states 
can be turned into an $N$-particle GHZ state by means of a suitable change of basis. In particular,
let us express $|B^{(N)}\rangle$ in terms of the ${\sigma}_x$ 
basis by writing for each spin 
$|\!\!\up\rangle= \frac{1}{\sqrt{2}}\, (|\!\leftarrow\rangle+|\!\rightarrow\rangle)$ and 
$|\!\!\dn\rangle= \frac{1}{\sqrt{2}}\, (|\!\leftarrow\rangle-|\!\rightarrow\rangle)$. Then we can show that 
$|B^{(N)}\rangle$ can be expressed in the form
\begin{eqnarray}
|B^{(N)}\rangle = \frac{1}{\sqrt{2}} \Big( |\!\leftarrow\leftarrow\cdots\leftarrow\rangle - 
|\!\rightarrow\rightarrow\cdots\rightarrow\rangle \Big) \, . 
\label{eq:12}
\end{eqnarray}
We can show then, after a short calculation, that if one traces out  any one of the particles, the resulting 
$(N-1)$-particle reduced  density matrix is separable, which establishes 
the claim for general $N$. 

To show that the $N$-particle ${\rm GHZ}_x$ state (\ref{eq:12}) emerges from the special Bernstein states, 
let us substitute $|\!\leftarrow \rangle = \frac{1}{\sqrt{2}}\, ( |\!\!\up\rangle + |\!\!\dn\rangle) $ and 
$|\!\rightarrow \rangle = \frac{1}{\sqrt{2}}\, ( |\!\!\up\rangle - |\!\!\dn\rangle)$ in (\ref{eq:12}). It is  
evident, due to the minus sign in (\ref{eq:12}), that $|B^{(N)}\rangle$ can be expressed as 
a linear superposition of all the $2^{N-1}$ states containing odd numbers of $\sigma_z$-down 
states. With this in mind, we can ask what  the probability is for a given particle being in the 
$\sigma_z$-up state. One only needs to count the number of terms in the expansion thus 
obtained for $|B^{(N)}\rangle$ that have the given particle in the $\sigma_z$-up state. But the 
number of terms for which an odd number of the remaining $N-1$ particles point in the 
$\sigma_z$-down direction is $2^{N-2}$. It follows that the probability of observing any given 
particle in the $\sigma_z$-up state is $\frac{1}{2}$, as required. Similarly, the probability of 
observing any given pair of particles in $\sigma_z$-up states can be obtained by noting that each 
relevant term in the expansion of $|B^{(N)}\rangle$ corresponds to a partition of the 
remaining $N-2$ particles into subsets of odd elements. There are $2^{N-3}$ such partitions, 
so the corresponding probability is $\frac{1}{4}$. Continuing in the same vein, one 
sees that the probability that any $k \leq N-1$ given particles are in the $\sigma_z$-up state is 
$2^{-k}$. However, not all of the $N$ particles can occupy $\sigma_z$-up states 
simultaneously, since the term proportional to $|\! \! \up \up \cdots \up \rangle$ is absent 
from the superposition making up $|B^{(N)} \rangle$. This establishes that, starting from 
the ${\rm GHZ}_x$ state (\ref{eq:12}), if we perform the reverse change of basis 
states for each spin, then the resulting state is the state 
that we obtained from the construction preceding (\ref{eq:11}). 

The foregoing analysis leads to a comment on the following rather counterintuitive property of GHZ states.  Suppose that we prepare an $N$-particle $ {\rm GHZ}_z$ state,
given by
\begin{eqnarray}
| \, {\rm GHZ}_z \, \rangle = \frac{1}{\sqrt{2}} \Big( |\!\up\up\cdots\up\rangle - 
|\!\dn\dn\cdots\dn\rangle \Big) .
\label{GHZ}
\end{eqnarray}
Then we enquire about the statistics arising from a projective measurement $\pi^{+}_{nx}$ 
of the spin for each particle in the $x$ direction. 
Since irrespective of 
whether the spin is in the $z$-up state or the $z$-down state, the probability of observing the 
$x$-up state is equal, one might expect that the probability of an affirmative outcome 
should be $\frac{1}{2}$. This intuition indeed is correct. Next we ask what statistics arise for 
joint projective measurements of this type for pairs of particles. By continuation of the 
intuitive argument, and taking into account the fact that the state $|\,{\rm GHZ}_z \rangle$ is totally 
symmetric, we conclude intuitively that
the probability of an affirmative outcome should be $\frac{1}{4}$; and this indeed 
turns out to be correct. However, when this  logic is extended all the way to the $N$ 
particle statistics, suddenly we find that intuition fails: the probability of an affirmative outcome  
is null, rather than $2^{-N}$. This probably comes as a surprise. Of course, if one were to 
perform the calculation by changing the basis in accordance with the above prescription, then 
one obtains a state of the form $|B^{(N)}\rangle$, from which the correct 
statistics can be deduced. However, if one does not perform this analysis, the conclusion is not perhaps
an obvious one.

\section{Construction of general Bernstein states} 
\label{sec:5} 

The special Bernstein states $|B^{(N)}\rangle$, $N=3, 4, \dots$, considered in the previous sections  
have been obtained under the assumption that the expansion coefficients are equal. What happens if we relax this 
condition? We know already that the inclusion of a phase factor in front of each of the basis states does not 
affect the Bernstein distribution. This follows from the fact that the  basis states are orthogonal to one another, and that when the  probabilities are calculated the phases cancel term by term. What is perhaps less obvious is that the squared amplitudes have to be the same. 
We proceed to prove that this is the case, and then 
turn to a discussion of the phase factors. 

To show that the magnitudes must be the same, we can use an inductive argument. Recall that for a system of 
$N$ spin-$\frac{1}{2}$ particles there are $2^N$ basis 
states that can be constructed from the single-particle $\sigma_z$ eigenstates.   
For the Bernstein distribution to hold,  the all-up basis state $|\!\up\up\cdots\up \rangle$ must have coefficient zero. What about 
the basis states with one spin pointing down---for example,  $|\!\up\dn\up\cdots\up \rangle$\,? 
Clearly, there are $N$ such states, one for each particle. The Bernstein distribution 
requires that the probability of finding any combination of $N-1$ spins in the up direction should
be $2^{1-N}$. Hence, the coefficients of each of these $N$ terms (up to phases) must be equal 
and given by $2^{(1-N)/2}$. Next, the contribution from these $N$ terms towards the 
probability of finding $N-2$ spins pointing in the up direction is clearly $2\times2^{1-N}=
2^{(2-N)}$, since for each fixed $N-2$ spins pointing in the up direction there are two 
possibilities for the remaining two spins (for example, $|\!\up\dn\up\cdots\up \rangle$ and 
$|\!\dn\up\up\cdots\up \rangle$ when we fix the last $N-2$ spins). 
But for the Bernstein distribution to hold we require that the probability of finding $N-2$ spins pointing in the up 
direction must be $2^{2-N}$, which we already have achieved from the $N$ terms, so there 
cannot be any other term containing $N-2$ spins in the up direction. It follows that the 
coefficients in front of the states containing two spin-down states (for example, 
$|\!\dn\dn\up\cdots\up \rangle$) must be zero. Continuing with the analysis of these $N$ terms, 
their contribution to the probability of finding $N-3$ particles in the up direction is clearly $3\times2^{1-N}$ (for 
example $|\!\up\up\dn\up\cdots\up \rangle$, $|\!\up\dn\up\up\cdots\up \rangle$, and 
$|\!\dn\up\up\up\cdots\up \rangle$ all contribute towards the last $N-3$ spins pointing in the 
up direction). The Bernstein distribution requires this probability to be $2^{3-N}$, so now we 
are short of terms. This can only be fixed by adding, for each fixed $N-3$ spins in the up direction, 
a term for which the remaining three spins are pointing in the down direction (in the previous 
example, we need to add $|\!\dn\dn\dn\up\cdots\up \rangle$). The coefficients of these 
three-spin-down states must all, up to phases, be given by $2^{(1-N)/2}$, since $2^{3-N}-3\times 
2^{1-N}=2^{1-N}$. 

We have shown that the coefficient of each term with a single spin pointing down or three spins pointing down is $2^{(1-N)/2}$, whereas all terms with no spins pointing down or two spins pointing down must be absent. 
The contribution from these existing terms towards the probability of finding $N-4$ spins pointing 
up  is given by $4\times2^{1-N}+4\times2^{1-N}=2^{4-N}$, which is the 
number required by the Bernstein distribution. It follows that the terms with four spins pointing  
down must be absent. Continuing inductively, one sees that all terms containing even numbers of spins pointing down  must be 
absent, and all terms containing odd numbers of spins pointing down must be 
present, with equal coefficients, up to phases. Hence, the construction 
of $|\,B^{(N)}\,\rangle$ presented in Section \ref{sec:3} gives the general form of a Bernstein state, up to phases. 

\section{Manifold of general Bernstein states} 
\label{sec:6} 
The foregoing analysis shows that the manifold ${\mathcal B}_{N}$ of projective $N$-particle Bernstein states is a torus $T^{2^{N-1}-1}$ in the
projective Hilbert space $\mathbb{PH}^{2^N-1}$ of pure $N$-particle states. More precisely, since there are ${2^{N-1}}$ terms in the basis expansion of a Bernstein state in the Hilbert space $\mathbb{H}^{2^N}$ of $N$ spin-$\frac{1}{2}$ particles, there are ${2^{N-1}}$ 
phase factors that do not affect the Bernstein probability distribution. The associated  ${2^{N-1}-1}$ relative phases parametrize a torus in $\mathbb{PH}^{2^N-1}$. 
An interesting question to ask is whether local  
unitary transformations of the pure phase type ('local phase transformations') are transitive on 
${\mathcal B}_{N}$. Equivalently, starting with a general Bernstein state, can the phase factors be eliminated 
by the application of a local phase transformation in such a way as to transform it into a special Bernstein state of the type 
constructed in Section~\ref{sec:3}, with equal real coefficients? The answer turns out to be no, except in the case $N=3$.  

This can be seen as follows. We fix a point $p$ in ${\mathcal B}_{N}$ and apply a local phase transformation  to it to obtain a new point ${\rm T} p$ in ${\mathcal B}_{N}$, which we call the transformed point. Thus, in each term of the expansion of the state vector representing the point $p$, we apply the phase factor 
$\exp({{\rm i}\alpha_k})$ if the state of particle $k$  is $|\!\!\up\rangle$, and we apply the phase factor 
$\exp({{\rm i}\beta_k})$ if the state of particle $k$ is $|\!\!\dn\rangle$. The components of the transformed state vector can be regarded as a set of homogeneous coordinates for ${\rm T} p$. In each term of the basis expansion of the transformed state vector, the phase of the corresponding term in the original state vector is shifted by a  linear combination of the various $\alpha_k$ and $\beta_k$. We call this combination the 'phase shift' of that term of the basis expansion. 

In the calculations below, it is convenient to remove a certain overall phase factor from the state vector in a way that ensures that the remaining coefficients depend on the various $\alpha_k$ and $\beta_k$ only through the $N$ phase differences defined by $\delta_k = \alpha_k-\beta_k$ for $k = 1, 2, \dots, N$.
For odd $N$ we remove an overall phase factor of the form $\exp\,(\sum_{k=1}^N\beta_k)$. For even $N$, we
remove an overall phase factor of the form $\exp\,(\alpha_N+\sum_{k=1}^{N-1}\beta_k)$. In both cases, once the relevant overall phase factor has been removed, we redefine the homogeneous coordinates of the transformed point ${\rm T} p$  accordingly. The new homogeneous coordinates for 
${\rm T} p$ clearly belong to the same projective equivalence class as the original homogeneous coordinates for ${\rm T} p$. It follows that the submanifold 
${\mathcal B}_N(p) \subset {\mathcal B}_{N}$ of projective Bernstein states that are attainable, starting at a given point $p \in {\mathcal B}_{N}$, 
by the action of local phase transformations is parametrized by a family of $N$ phase 
differences, namely the $\delta_k$ defined above.

We shall show that for each choice of $p$ there is a one-to-one invertible map from the points of an $N$-torus to the 
points of ${\mathcal B}_N(p)$. Consider the $N$ terms in the expansion of 
the transformed Bernstein state that contain a single spin-down particle. Denoting 
the phase shifts appearing in these terms by $\phi_k$ where $k =1 ,2,  \ldots, N$, we find that the various  $\phi_k$ can be expressed in terms of the various  $\delta_k$ by the relation
\begin{eqnarray}
\left( \begin{array}{ccccc} 
1 & 1 & \cdots & 1 & 0 \\ 
1 & \vdots & \revddots & 0 & 1 \\ 
\vdots & 1 & \revddots & 1 & \vdots \\ 
1 & 0 & \revddots & \vdots & 1 \\
0 & 1 & \cdots & 1 & 1 
\end{array} \right) \left( \begin{array}{c} 
\delta_1 \\ \delta_2 \\ \vdots \\ \delta_{N-1} \\ \delta_N \end{array} \right) = 
\left( \begin{array}{c} 
\phi_1 \\ \phi_2 \\ \vdots \\ \phi_{N-1} \\ \phi_N \end{array} \right)
\label{eq:16} 
\end{eqnarray}  
when $N$ is odd, since for each term with one of the spins pointing down, the phase shift is given by a sum of all but 
one of the $\delta_k$. Similarly, when $N$ is even we obtain 
\begin{eqnarray}
\left( \begin{array}{ccccc} 
1 & 1 & \cdots & 1 & -1 \\ 
1 & \vdots & \revddots & 0 & 0 \\ 
\vdots & 1 & \revddots & 1 & \vdots \\ 
1 & 0 & \revddots & \vdots & 0 \\
0 & 1 & \cdots & 1 & 0 
\end{array} \right) \left( \begin{array}{c} 
\delta_1 \\ \delta_2 \\ \vdots \\ \delta_{N-1} \\ \delta_N \end{array} \right) = 
\left( \begin{array}{c} 
\phi_1 \\ \phi_2 \\ \vdots \\ \phi_{N-1} \\ \phi_N \end{array} \right) .
\label{eq:16.2} 
\end{eqnarray} 
Note that the matrix of coefficients in (\ref{eq:16.2}) is obtained by subtracting unity from each entry of the last column 
of the matrix of coefficients in (\ref{eq:16}). For both $N$ even and $N$ odd, the relevant matrix of coefficients is invertible, 
and one finds that  
\begin{eqnarray}
\left( \begin{array}{c}
\delta_1 \\ \delta_2 \\ \vdots \\ \delta_{N-1} \\ \delta_N \end{array} \right) =
\frac{1}{N-1}
\left( \begin{array}{ccccc}
1 & 1 & \cdots & 1 & 2-N \\
1 & \vdots& \revddots & 2-N & 1 \\
\vdots & 1 & \revddots & 1 & \vdots \\
1 & 2-N & \revddots & \vdots & 1 \\
2-N & 1 & \cdots & 1 & 1 
\end{array} \right)
\left( \begin{array}{c}
\phi_1 \\ \phi_2 \\ \vdots \\ \phi_{N-1} \\ \phi_N \end{array} \right)
\label{eq:17}
\end{eqnarray}
if $N$ is odd, and that
\begin{eqnarray}
\left( \begin{array}{c}
\delta_1 \\ \delta_2 \\ \vdots \\ \delta_{N-1} \\ \delta_N \end{array} \right) =
\frac{1}{N-2}
\left( \begin{array}{ccccc}
0 & 1 & \cdots & 1 & 3-N \\
0 & \vdots & \revddots & 3-N & 1 \\
\vdots & 1 & \revddots & 1 & \vdots \\
0 & 3-N & \revddots & \vdots & 1 \\
2-N & 1 & \cdots & 1 & 1
\end{array} \right)
\left( \begin{array}{c}
\phi_1 \\ \phi_2 \\ \vdots \\ \phi_{N-1} \\ \phi_N \end{array} \right)
\label{eq:17.2}
\end{eqnarray}
if $N$ is even. 
Thus, the various  $\delta_k$ can be expressed in terms of the various  $\phi_k$. 

The point $p$ is left invariant under the given local phase transformation if and only if the phase shifts
are integral multiples of $2\pi$. More generally, two distinct local unitary transformations of the pure phase type transform $p$ to the same point  if and only if the associated $\phi_k$ 
differ by integral multiples of $2\pi$. This allows us to identify a toric periodicity structure 
in the space of $\delta_k$, which can be read off from the column vectors of the 
matrix in (\ref{eq:17}) when $N$ is odd and from  the column vectors of the 
matrix (\ref{eq:17.2}) when $N$ 
is even. For instance, in the case $N=3$ we see that (\ref{eq:17}) implies that 
\begin{eqnarray}
\left( \begin{array}{c} 
\delta_1 \\ \delta_2 \\ \delta_3 \end{array} \right) = 
\frac{1}{2} 
\left( \begin{array}{cccc} 
1 & 1 & -1 \\ 
1 & -1 & 1 \\ 
-1 & 1 & 1  
\end{array} \right) 
\left( \begin{array}{c} 
\phi_1 \\ \phi_2 \\ \phi_3 \end{array} \right).
\label{eq:18} 
\end{eqnarray}  
It follows that $p$ is invariant under precisely those transformations that satisfy
\begin{equation}
  \left( \begin{array}{c} 
\delta_1 \\ \delta_2 \\ \delta_3 \end{array} \right) = l \pi \left( \begin{array}{c} 
1 \\ 1 \\ -1 \end{array} \right) + m \pi  \left( \begin{array}{c} 
1 \\ -1 \\ 1 \end{array} \right) + n \pi \left( \begin{array}{c} 
-1 \\ 1 \\ 1 \end{array} \right) 
\label{eq:18_dash}
\end{equation}
for some collection of integers $l, m, n$. More generally, we see that 
$\delta_k$ and $\bar \delta_k$ transform $p$ to the same point if and only if 
\begin{equation}
  \left( \begin{array}{c} \bar
\delta_1 \\ \bar \delta_2 \\ \bar \delta_3 \end{array} \right) =
  \left( \begin{array}{c} 
\delta_1 \\ \delta_2 \\ \delta_3 \end{array} \right) +
 l \pi \left( \begin{array}{c} 
1 \\ 1 \\ -1 \end{array} \right) + m \pi  \left( \begin{array}{c} 
1 \\ -1 \\ 1 \end{array} \right) + n \pi \left( \begin{array}{c} 
-1 \\ 1 \\ 1 \end{array} \right). 
\label{eq:18_dashdash}
\end{equation}
Once one has made the appropriate identifications, the points of the resulting torus are in one-to-one correspondence with the  states that can be reached from the given Bernstein state $p$ by use of local phase
transformations. 

In the case $N=4$ 
we have 
\begin{eqnarray}
\left( \begin{array}{c} 
\delta_1 \\ \delta_2 \\ \delta_3 \\ \delta_4 \end{array} \right) =\frac{1}{2}\left( 
\begin{array}{cccc} 
0 & 1 & 1 & -1 \\ 
0 & 1 & -1 & 1 \\ 
0 & -1 & 1 & 1 \\ 
-2 & 1 & 1 & 1 
\end{array} \right) \left( \begin{array}{c} 
\phi_1 \\ \phi_2 \\ \phi_3 \\ \phi_4 \end{array} \right).
\end{eqnarray} 
This implies that a given Bernstein state $p$ is  left invariant by precisely those transformations for which 
\begin{equation}
\left( \begin{array}{c} 
\delta_1 \\ \delta_2 \\ \delta_3 \\ \delta_4 \end{array} \right) = 
k\pi \left( \begin{array}{c} 
0\\ 0 \\ 0 \\ -2 \end{array} \right) + l \pi \left( \begin{array}{c} 
1 \\ 1 \\ -1 \\ 1 \end{array} \right) + m \pi \left( \begin{array}{c} 
1 \\ -1 \\ 1 \\ 1 \end{array} \right) + n \pi \left( \begin{array}{c} 
-1 \\ 1 \\1 \\ 1 \end{array} \right)
\label{eq:x23} 
\end{equation}
for some collection of integers $k$, $l$, $m$, and $n$. More generally, $\delta_k$ and $\bar \delta_k$ transform $p$ to the same point if and only if 
\begin{equation}
\left( \begin{array}{c} \bar
\delta_1\\ \bar \delta_2 \\  \bar \delta_3 \\ \bar  \delta_4 \end{array} \right)=\left( \begin{array}{c} 
\delta_1 \\ \delta_2 \\ \delta_3 \\ \delta_4 \end{array} \right) +
k\pi \left( \begin{array}{c} 
0\\ 0 \\ 0 \\ -2 \end{array} \right) + l \pi \left( \begin{array}{c} 
1 \\ 1 \\ -1 \\ 1 \end{array} \right) + m \pi \left( \begin{array}{c} 
1 \\ -1 \\ 1 \\ 1 \end{array} \right) + n \pi \left( \begin{array}{c} 
-1 \\ 1 \\1 \\ 1 \end{array} \right).
\label{eq:x23_dash} 
\end{equation}
Equation (\ref{eq:x23_dash}) allows one to identify the periods of a  torus, the points of which are in one-to-one correspondence with the states attainable from the given Bernstein state $p$ by use of local phase transformations.

More generally, the submanifold ${\mathcal B}_N(p)$ of Bernstein states that can be reached from an initial state $p$  
by local phase transformations takes the form of a nonsingular embedding of an $N$-torus 
in ${\mathcal B}_N$,  with a periodicity structure similar to
(\ref{eq:18_dashdash}), (\ref{eq:x23_dash}). For  $N>3$, this $N$-torus is of lower dimension than the Bernstein 
manifold ${\mathcal B}_N$, which is a torus of dimension $2^{N-1}-1$. Thus, for $N>3$ the phase factors cannot be eliminated by local phase
transformations. But for $N=3$ we have 
$2^{3-1}-1=3$.

An implication of this observation is that the ${\rm GHZ}_x$ states, together with those states accessible from a 
${\rm GHZ}_x$ state via local phase transformations, form a proper subset of the space of general Bernstein 
states---the only exception being the case $N=3$, for which ${\rm GHZ}_x$ states and general Bernstein states are 
equivalent modulo local phase  transformations. This subset is of particular interest since all its elements satisfy the fragility property, namely,   if one traces out the degree of freedom associated with any one of the 
particles, the resulting density matrix is separable. 
  
\section{Inhomogeneous Bernstein states} 
\label{sec:7}

In the formulation of the quantum Bernstein distribution that we have considered so far, we have required that the relevant probabilities should be symmetric in 
the sense that the probability of any one given particle being in a spin-up state should be $\frac{1}{2}$, the probability of any two given particles being in up states should be $\frac{1}{4}$, and so on, all the way up to $N-1$ particles. The resulting states are unique up to phase factors. 

If we relax the symmetry property, and merely require that the measurement outcomes
for any choice of $2, 3, \ldots, N-1$ particles are independent, and that the measurement outcomes for all
$N$ particles 
are not independent, then there are many more possibilities. We shall refer to such states as 
an inhomogeneous Bernstein states. As an example, consider the state 
\begin{eqnarray}
|{B}_q^{(3)} \rangle &=& 
q\Big( |\! \uparrow\uparrow\downarrow\rangle + |\! \uparrow\downarrow\uparrow\rangle 
+ |\! \downarrow\uparrow\uparrow\rangle \Big) \nonumber \\ && + \sqrt{q(1-2q)} \, \Big( 
|\! \uparrow\downarrow\downarrow\rangle + |\! \downarrow\uparrow\downarrow\rangle + 
|\! \downarrow\downarrow\uparrow\rangle \Big) + \sqrt{1-3q(1-q)} \, 
|\! \downarrow\downarrow\downarrow\rangle 
\end{eqnarray}
for $0<q\leq\frac{1}{2}$. For $q=\frac{1}{2}$ we have $|B_{\frac{1}{2}}^{\,(3)}\rangle=|B^{(3)}\rangle$. 
For general $q$ we find that  
\begin{eqnarray} 
{\mathbb P}(+\bullet\bullet)={\mathbb P}(\bullet+\bullet)={\mathbb P}(\bullet\bullet+)=q \, .
\end{eqnarray} 
for measurements on a single particle, and that 
\begin{eqnarray} 
{\mathbb P}(++\bullet)={\mathbb P}(+\bullet+)={\mathbb P}(\bullet++)=q^2
\end{eqnarray}
for measurements on pairs of particles. 
Hence the outcomes are statistically independent. Yet,  for measurements on three particles we have
\begin{eqnarray} 
{\mathbb P}(+++)= 0 \neq 
{\mathbb P}(+\bullet\bullet)\times{\mathbb P}(\bullet+\bullet)\times{\mathbb P}(\bullet\bullet+)
=q^3.
\end{eqnarray}
Thus we have a generalization of the Bernstein statistics. 

On the other hand, the reduced density matrix 
${\rm tr}_{{\rm P}_1}(|{B}_q^{(3)}\rangle\langle {B}_q^{(3)}|)$ that results if we trace out the first particle is not separable unless 
$q=\frac{1}{2}$, and the same holds for the other two reduced density matrices.  Hence 
the inhomogeneous Bernstein states do not inherit the fragility property. 

\section{Remarks on Mermin's paradox} 
\label{sec:8} 

The analysis that we have undertaken in the previous sections can be used to gain some insight into aspects of the so-called Mermin paradox \cite{mermin}. In particular, with a slight modification of our earlier calculations one can obtain a generalization of Mermin's result to any number of particles. 

It will be convenient to begin by
reviewing the ideas of \cite{mermin} in a version presented by Peres \cite{peres}. A standard three-particle ${\rm GHZ}_z$ state of the form (\ref{GHZ}) 
is prepared, and one considers the outcomes of product spin measurements made using
four different combinations of directions, given by (x,y,y), (y,x,y), (y,y,x), and (x,x,x). The 
observable corresponding to the (x,y,y) combination, for instance, is given by 
${\sigma}_{1x}\otimes  {\sigma}_{2y}\otimes {\sigma}_{3y}$.  The GHZ state is a simultaneous eigenstate of these 
four product observables. To see that this is the case, we recall that  
\begin{eqnarray}
|\!\!\uparrow\rangle = \frac{1}{\sqrt{2}}(|\!\!\leftarrow\rangle + |\!\!\rightarrow\rangle)  
=  \frac{1}{\sqrt{2}} (|\!\!\nearrow\rangle + {\rm i} |\!\!\swarrow\rangle) 
\end{eqnarray} 
and 
\begin{eqnarray}  
|\!\!\downarrow\rangle =  \frac{1}{\sqrt{2}} (|\!\!\leftarrow\rangle - |\!\!\rightarrow\rangle)  =  \frac{-{\rm i}}{\sqrt{2}} ( |\!\!\nearrow\rangle - {\rm i} |\!\!\swarrow\rangle) ,  
\end{eqnarray}
where we write $|\!\!\nearrow\rangle$ and $|\!\!\swarrow\rangle$ for the $y$-up and the 
$y$-down states. To calculate the eigenvalue of the ${\rm GHZ}_z$ state for the (x,y,y) observable, one expands the state of the first particle in the $x$-basis and the states of the remaining particles in the $y$-basis. Expressed in the new basis states, only terms with an even number of downward spins remain.  Hence, the eigenvalue of ${\sigma}_{1x}\otimes  {\sigma}_{2y} \otimes {\sigma}_{3y}$ is $1$. Analogous calculations for 
(y,x,y) and (y,y,x) show that the relevant eigenvalues take the value $1$. To compute the eigenvalue for (x,x,x), one expands all the particles  in the $x$ basis. Only terms with an odd number of downward spins appear,  leading to an eigenvalue of $-1$. 
A paradox then arises if one assumes that preassigned outcomes can be assigned to the various individual spin measurements, depending on which direction is chosen for the measurement axis. If we write $m_{nx}$ for the outcomes of the 
${\sigma}_{nx}$ measurements, and $m_{ny}$ for the outcomes of the 
${\sigma}_{ny}$ measurements, then it follows that 
$m_{1x}m_{2y}m_{3y}=1$, $m_{1y}m_{2x}m_{3y}=1$, $m_{1y}m_{2y}m_{3x}=1$, and $m_{1x}m_{2x}m_{3x}=-1$. On account of the fact that $m_{nx}^2=m_{ny}^2=1$, 
the product of the left sides of these four equations is $+1$, whereas the product of the 
right sides is $-1$, and hence we have a contradiction.  

In the calculations of the eigenvalues above we made use of the method of particle-wise 
coordinate rotation, examples of which we have already encountered. For instance, (\ref{eq:1}), (\ref{eq:6}) and (\ref{eq:11}) are basis-transformed expressions of the ${\rm GHZ}_x$ states for three, four and five particles, respectively.  It turns out that an adaptation of the $N$-particle version of this method leads to a generalization of \cite{mermin} for $N > 3$. Consider an observable consisting of $2k$ factors of $ {\sigma}_y$ and $N-2k$ factors of  $ {\sigma}_x$. By expressing the respective $2k$ spin vectors in $y$-basis and the remaining $N-2k$ spin vectors in the $x$-basis, one observes two patterns. If $k$ is even, then only terms with an overall odd number of down spins remain. If $k$ is odd, then all terms contain an even number of down spins. As before, this is inconsistent with preassigned values. 

For example, if $N = 4$ then two of the relevant observables are given by the 
combinations (x,x,x,x) and 
(y,y,y,y). Here, $k = 0$ and $k = 2$ respectively, hence an odd overall number of downward spins remain, corresponding to an eigenvalue of $-1$. There are six additional observables, each with two factors of ${\sigma}_y$. For these we have $k = 1$, and hence the eigenvalue is $1$. We therefore obtain the relations
\begin{eqnarray} 
m_{1x}m_{2x}m_{3x}m_{4x}=-1, \quad  m_{1y}m_{2y}m_{3y}m_{4y}=-1 
\label{merm4eq:1}
\end{eqnarray} 
and
\begin{eqnarray} 
&&   m_{1x}m_{2x}m_{3y}m_{4y}=1, \quad   m_{1x}m_{2y}m_{3x}m_{4y}=1, \quad   m_{1x}m_{2y}m_{3y}m_{4x}=1, \nonumber \\ &&   m_{1y}m_{2x}m_{3x}m_{4y}=1, \quad   m_{1y}m_{2x}m_{3y}m_{4x}=1, \quad   m_{1y}m_{2y}m_{3x}m_{4x}=1. \label{merm4eq:2}
\end{eqnarray}
 Unlike in the $N=3$ case, where one 
encounters a single contradiction, now we encounter several distinct quadruplets of equations leading to contradictions. One such quadruplet, for instance, consists of the first equation of (\ref{merm4eq:1}), with equations one, two and three of (\ref{merm4eq:2}). One can replace equations two and three by equations four and five to obtain a distinct quadruplet. In total, one finds \emph{eight} distinct quadruplets.

As another example, let us consider the case $N=5$. The ${\rm GHZ}_z$ state is a simultaneous eigenstate of  16 observables containing even factors of ${\sigma}_y$, namely, 1 of the form (x,x,x,x,x), 5 of the form (x,y,y,y,y), and 10 of the 
form (x,x,x,y,y), the last set of 10 having eigenvalue $1$ and all others $-1$. That is,
\begin{eqnarray} 
\qquad && \fl  m_{1x}m_{2x}m_{3x}m_{4x}m_{5x}=-1 
\label{eq:18.2}
\end{eqnarray} 
for the zero-$y$ measurements, 
\begin{eqnarray} 
\qquad && \fl  m_{1x}m_{2y}m_{3y}m_{4y}m_{5y}=-1, \quad m_{1y}m_{2x}m_{3y}m_{4y}m_{5y}=-1, \quad 
m_{1y}m_{2y}m_{3x}m_{4y}m_{5y}=-1, \nonumber \\  \qquad && \fl m_{1y}m_{2y}m_{3y}m_{4x}m_{5y}=-1, 
\quad m_{1y}m_{2y}m_{3y}m_{4y}m_{5x}=-1
\label{eq:19}
\end{eqnarray} 
for the four-$y$ measurements, and 
\begin{eqnarray} 
\qquad && \fl  m_{1x}m_{2x}m_{3x}m_{4y}m_{5y}=1, \quad m_{1x}m_{2x}m_{3y}m_{4x}m_{5y}=1, \quad 
m_{1x}m_{2y}m_{3x}m_{4x}m_{5y}=1, \nonumber \\ \qquad  && \fl m_{1y}m_{2x}m_{3x}m_{4x}m_{5y}=1, 
\quad m_{1x}m_{2x}m_{3y}m_{4y}m_{5x}=1, \quad m_{1x}m_{2y}m_{3x}m_{4y}m_{5x}=1, 
\nonumber \\ \qquad && \fl m_{1y}m_{2x}m_{3x}m_{4y}m_{5x}=1, \quad m_{1x}m_{2y}m_{3y}m_{4x}m_{5x}=1, 
\quad m_{1y}m_{2x}m_{3y}m_{4x}m_{5x}=1, \nonumber \\ \qquad && \fl m_{1y}m_{2y}m_{3x}m_{4x}m_{5x}=1
\label{eq:20}
\end{eqnarray} 
for the two-$y$ measurements. Now we encounter many contradictions. For each equation in (\ref{eq:18.2}) or (\ref{eq:19}) one can find one or more 
combinations of three equations in (\ref{eq:20}) that lead to inconsistencies, and likewise for 
each equation in (\ref{eq:20}) one can find a combination of three equations in (\ref{eq:18.2}) or 
(\ref{eq:19}) that leads to an inconsistency. The number of such contradictions grows as 
$N$ is increased. 

It remains a challenge to find practical applications of the unusual properties of Bernstein 
states. Since some of the 
configurations that we have considered give rise to situations where a fully entangled $N$-particle state 
separates if any one of the particles is traced over, the results may find applications in quantum cryptography, for example, in the area of `secret sharing' 
protocols~\cite{kashefi}. An interesting application of the Bernstein 
distribution in connection with the so-called 'principle of common cause' has been discussed in \cite{Uffink}. 
The approach to experimental realizations of Bernstein-type statistics in the 
context of photon polarization described in \cite{Belinsky,Chirkin} offers a further possible direction for practical developments.



\end{document}